\documentclass[preprint,12pt,3p]{elsarticle}
\usepackage[english]{babel}
\usepackage{times}
\usepackage{hyphenat}
\usepackage{booktabs}
\usepackage{gensymb}
\usepackage{siunitx}
\usepackage{amsmath}
\usepackage{subfig}
\usepackage[figuresright]{rotating}
\usepackage{multirow}
\usepackage{setspace}
\usepackage{hyperref}
\usepackage[nomarkers]{endfloat}
\usepackage{lipsum}
\usepackage{everysel}
\usepackage{lineno,xcolor}
\usepackage{wrapfig}
\usepackage{adjustbox}
\biboptions{sort&compress}
\usepackage[cmintegrals,cmbraces]{newtxmath}
\usepackage[utf8]{inputenc} % usually not needed (loaded by default)
\usepackage[superscript]{cite}

\usepackage[sfdefault,condensed]{roboto}  %% Option 'sfdefault' only if the base font of the document is to be sans serif
\usepackage[T1]{fontenc}

\begin{document}
	
	\begin{frontmatter}
		
		\journal{xxxx}
		
		\title{\Large\textbf{A New Methodology for Radiation Effects Studies in Solids using the Plasma Focused Ion Beam}}
		
		\author[LANL]{M.A. Tunes\corref{cor}}\ead{tunes@lanl.gov}\ead[url]{materialsatextremes.wordpress.com}
		\cortext[cor]{Corresponding author.}
		\author[LANL]{M.M. Schneider}
		\author[LANL]{C.A. Taylor}
		%\author[UOM]{P.D. Edmondson}
		\author[LANL]{T.A. Saleh}
		
		\address[LANL]{Materials Science and Technology Division, Los Alamos National Laboratory, United States of America}
		%\address[UOM]{Department of Materials, The University of Manchester, United Kindgom}

		\begin{abstract}
			\onehalfspacing
			
			\noindent A new methodology for fundamental studies of radiation effects in solids is herein introduced by using a plasma Focused Ion Beam (PFIB). The classical example of ion-induced amorphization of single-crystalline pure Si is used as a proof-of-concept experiment that delineates the advantages and limitations of this new technique. We demonstrate both the feasibility and invention of a new ion irradiation mode consisting of irradiating a single-specimen in multiple areas, at multiple doses, in specific sites. This present methodology suggests a very precise control of the ion beam over the specimen, with an error in the flux on the order of only 1\%. In addition, the proposed methodology allows the irradiation of specimens with higher dose rates when compared with conventional ion accelerators and implanters. This methodology is expected to open new research frontiers beyond the scope of materials at extremes such as in nanopatterning and nanodevices fabrication.
			
		\end{abstract}
		
		
	\end{frontmatter}
	\onehalfspacing
	
	\section*{Introduction}
	\label{sec:introduction}
	
	\noindent The intense pace of technological developments in our modern and digitally connected society poses new demands for a significant increase in electric power energy generation. Concurrently, energy must now be generated in a cleaner way in order to meet the desired goals for both economic sustainability and carbon neutrality. In the context of these worldwide efforts, a series of new energy sources are now under intense research and development stages aiming at future commercialization \cite{lewis2016paris}. 
	
	It is widely known that nuclear energy is a carbon-free source of power and is currently experiencing a renaissance period \cite{cattant2008corrosion}. Among several different new ideas proposed for the energy transition of our time \cite{kittner2017energy}, new designs for fission- and fusion-based nuclear reactors are envisaged to replace the large scale use of fossil-based fuels and their derivatives. These reactors are theoretically capable of generating the highest energy densities (energy release per unit of fuel mass) ever achieved by mankind in an inexhaustible manner \cite{ongena2004energy}, but both development and commissioning stages still require the invention of novel materials capable of resisting their inherent harsh environmental conditions \cite{yvon2009structural,allen2010materials,cesario2010current,zinkle2013materials,zhang2015ionization}. The triad of materials degradation mechanisms present in the extreme environment of prototypic fission and fusion reactors are: corrosion, high-temperature, and energetic particle irradiation. Therefore, new candidate materials to fit these new forms of power are required to exhibit both high microstructural phase stability and mitigation of atomic- and nano-scale defect generation when subjected to the operation of such degradation mechanisms. These new guidelines for materials design and selection may guide the scientific research within metallurgy and materials science for the next decades.
	
	In order to be considered for application as a structural material in a future nuclear reactor design, a material must be exhaustively tested and evaluated under the operation of these three degradation mechanisms to avoid the occurrence of materials related nuclear accidents. Nowadays, engineering protocols for corrosion and high-temperature testing are already considered fast and reliable within the nuclear industry \cite{cattant2008corrosion}, but the evaluation of potential nuclear materials can still take decades to complete. The latter fact is due to mandatory national and international regulations that require such new materials to be irradiated in Materials Test Reactors (MTRs) in order to probe their response to energetic particle irradiation exposure \cite{aguiar2020bringing}. 
	
	Two experimental methodologies are of historical importance when testing materials under irradiation : neutron irradiation using MTRs and ion irradiation using particle accelerators. It is both economically and scientifically impracticable to use MTRs to evaluate the entire spectrum of potential nuclear materials. Expensive tests on the order of millions of dollars, low displacement-per-atom (dpa) dose rate and yield, and the induced radioactive activation of specimens are factors to consider. Thus, the use of particle accelerators and ion beams arises as a rapid quality assessment option: if a candidate material performs well under specific ion irradiation conditions that roughly emulate neutron irradiation, then this material can be selected for further evaluation in MTRs. In fact, it has been known since late 1970s that the use of heavy-ions (preferably inert gases) with energies between 40 and 240 keV are a viable alternative methodology to emulate both the morphology and the kinetics of the displacement cascade of defects generated by neutron irradiation in most metals of interest to the nuclear industry \cite{english1976heavy}. However, as with any experimental technique, ion bombardment of solids has its limitations that often are not fully documented in literature. Some examples of such limitations are: asymmetries in the ion beam geometry resulting in inhomogeneous irradiation profiles, poor control and monitoring of the ion beam flux over time, limited flux and dose rates, limited accuracy in the estimation of the ion beam current on a target, low accuracy in establishing a precise area for irradiation on a target, and so on.
	
	Unquestionably, new faster and more reliable experimental techniques are required to foster discoveries of new materials for application in extreme environments. The use of the Plasma Focused Ion Beam (PFIB) as a new experimental technique to evaluate materials under irradiation is the main objective of this present work. Recently introduced as a new experimental capability for research within the field of materials science, the PFIB has the major advantage of featuring a Xe$^{+}$ ion gun as opposed to previous Ga$^{+}$ ion-based FIB systems. The PFIB acceleration energy is also limited to 30 keV, but the inert gas does not deleteriously interact with most metals as Ga sometimes does \cite{phaneuf2005fib,phaneuf2002gallium,bala2022effect}. In addition, the heavier inert gas ion allows for faster milling and trenching rates which accelerates all the intended applications of FIB systems. Historically, it has been highly desirable to avoid radiation-induced damage caused by the FIB when preparing samples for either Transmission Electron Microscopy (TEM) or micro-mechanical testing \cite{giannuzzi1999review,pfeifenberger2017use,mayer2007tem,tweddle2022direct}. Conversely, we herein intend to test our hypothesis that the PFIB can be used for controlled ion irradiation experiments. While several recent studies using 30 keV Xe$^{+}$ ion irradiation with \textit{in situ} TEM have demonstrated viability for using this ion energy to address the radiation response of conventional nuclear materials \cite{tunes2019radiation,tunes2019thermodynamics,tunes2021comparative}, the use of the PFIB as a methodology for radiation effects studies in solids was, until now, pending technique development.  
	
	To evaluate the feasibility of using the PFIB as a platform for controlled radiation effects studies, we have chosen a case study that has been thoroughly investigated over the course of the ion beam modification of materials history: the amorphization of pure single-crystal Si \cite{edmondson2009amorphization,edmondson2016situ}. As the basis of modern semiconductor microelectronics that use ion beams for doping \cite{leong1997silicon,prati2012anderson}, irradiation of pure and single-crystalline Si is known to degrade its electronic transport properties, induce ionization, and progressively destroy its long-range order (LRO) \cite{holmes1993handbook}.  In brief, ion impacts into Si generate thermal and displacement spikes \cite{brinkman1954nature} that cause a local melt of the material within the displacement cascade core. After spike relaxation, a localized amorphous pocket is created within the material. As the ion fluence increases, these amorphous pockets (or zones) interact and grow, resulting in loss of LRO and thus, amorphization \cite{edmondson2009amorphization,edmondson2016situ,larubia1996defect,howe1981features,howe1987heavy}. A theoretical refinement on the models for amorphization of semiconductors has been presented by Gibbons who indicated that an overlap of displacement cascades is necessary step to create such amorphous zones \cite{gibbons1972ion,weber2000models}. This is known as the Gibbons's heterogenous model. His ideas were recently largely validated with experiments using \textit{in situ} TEM light and heavy-ion irradiations \cite{edmondson2009amorphization,edmondson2016situ} as well as via 3D reconstruction of displacement cascades \cite{camara2019understanding}, both indicating the predominance of the heterogenous model over homogenous models that only considered build-up of point defects as the main cause for amorphization \cite{edmondson2009amorphization}. Notwithstanding, the experiments reported in this present research work serve as a ``proof-of-concept'' to assess whether the PFIB can be viewed as suitable new tool for the radiation effects community.
	
	\section*{Results}
	\label{sec:results}
	
	\subsection*{Radiation effects research using the PFIB}
	\label{sec:results:methodology}
	
	\noindent The developed methodology to allow ion beam irradiation using the PFIB  is shown in \textbf{Fig. \ref{fig:01}}. Under normal ``imaging'' conditions, the ion beam is focused to a fine point while it is rastered over the sample. The spacing between adjacent points in the raster pattern is defined by the magnification and pixel-resolution specified by the current imaging conditions. For irradiations, the raster pattern is modified and the spacing between adjacent irradiated points is defined to be 50\% of the ideal beam profile. This beam-placement spacing ensures that the entire region of interest is irradiated and there are no underexposed/unexposed areas. Subsequently, low dwell times are used to establish a \textit{quasi}-instantaneous ion beam exposure over time and onto the target irradiation area. This procedure is carried out both with the specimen at a normal incidence angle with respect to the ion column (conventionally set to 52$^{\circ}$) and by using the rectangle tool in the ion beam mode. In this way, irradiation current (flux), time and area can be accurately controlled. As a commercial scientific instrument, current measurements in the PFIB  have an error of <1\% (see details in Materials and Methods), which propagates to the flux. Irradiations with the 30 keV Xe$^{+}$ ion beam will always be limited by the thickness of the target as shown in the SRIM calculations presented in \textbf{Fig. \ref{fig:02}A-C}. Under these irradiation conditions on pure and electron-transparent Si (thickness $\approx$40 nm, as measured with EELS \cite{egerton2008electron}), the projected Xe implantation range in \textbf{Fig. \ref{fig:02}A} indicates that all Xe will be retained within the 40 nm thick specimen, whilst the damage profile in \textbf{Fig. \ref{fig:02}B} peaks at the first Si layers, then it decreases as a function of the specimen's thickness. Using a 25$\times$25 $\mu$m$^2$ irradiation area and an ion beam current of 30 pA, the irradiation flux is 3.0$\times$10$^{13}$ ions$\cdot$cm$^{-2}\cdot$s$^{-1}$ and the average dose rate was estimated to be 0.042 dpa$\cdot$s$^{-1}$ (see \textbf{Fig. \ref{fig:02}C}). Very high average doses can be reached in less than an hour in that specific area (100 dpa per 40 min). Xe fluxes at this order of magnitude are normally unattainable in conventional ion accelerators and implanters dedicated to the study of radiation effects in nuclear structural materials \cite{tunes2019radiation,tunes2019thermodynamics,tunes2021comparative}. Both methodology and SRIM calculations suggest that ion irradiation within the PFIB will be primarily applicable to electron-transparent TEM specimens. Its use for irradiations on bulk materials is pending further research investigation beyond the scope of the present work.
	
	\subsection*{A new multi-area, multi-dose and site-specific irradiation mode}
	\label{sec:results:controlled}
	
\begin{figure}	
		\centering	
		\includegraphics[width=\textwidth]{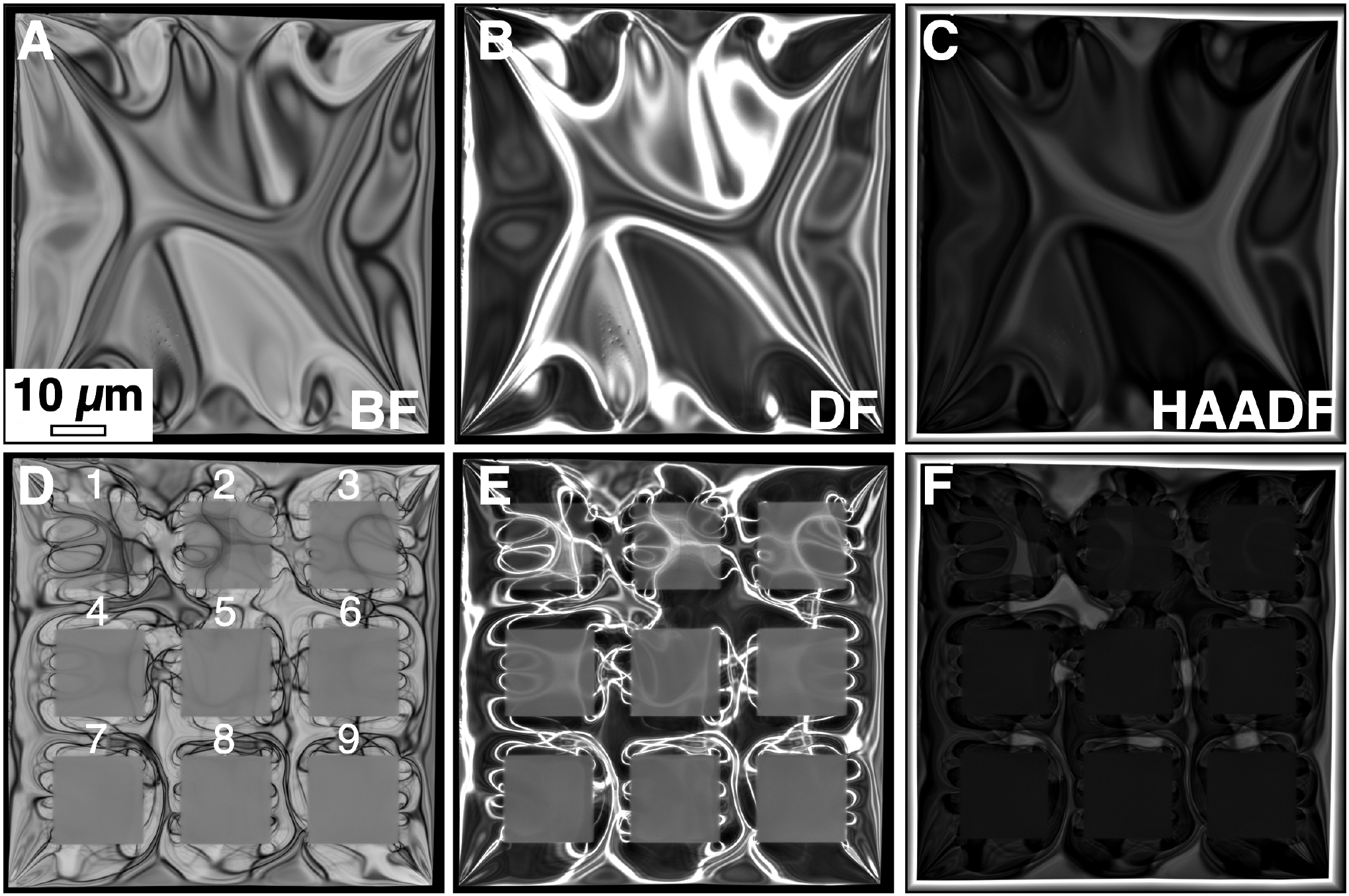}
		\caption{\textbf{Controlling amorphization of Si in a new multi-area, multi-dose and site-specific irradiation mode} | SEM-STEM micrographs \textbf{A, B} and \textbf{C} were taken before irradiation and exhibit the microstructure of the single-crystal pure Si specimen closer to the [001] zone axis with bright-field (BF), dark-field (DF) and high-angle annular dark-field detectors (HAADF), respectively. Then, sample was tilted towards the ion beam and a set of nine areas consisting of 15$\times$15 $\mu$m$^{2}$ each were drawn in the ion beam mode. Each area was set to a different irradiation time, starting with 1 until 9 s. The dose rate was estimated to be 0.035 dpa$\cdot$s$^{-1}$ with a set current of 10 pA for this experiment. Post-irradiation characterization in SEM-STEM mode is shown in micrographs \textbf{D, E} and \textbf{F} exhibiting BF, DF and HAADF, respectively. Upon losing the bending contour contrast as a function of the irradiation time, one can infer the amorphization of the pure Si specimen confined only in the nine squares. Outside the irradiated areas, the Si specimens remains fully crystalline as noted with the presence of strong bending contour contrast. This arises a new methodology for radiation effects investigations: multi-area,  multi-dose and site-specific irradiation mode in a single specimen. Note: scale bar in \textbf{A} applies to all micrographs in the figure.}
		\label{fig:03}
	\end{figure}

\begin{figure}	
		\centering	
		\includegraphics[width=\textwidth]{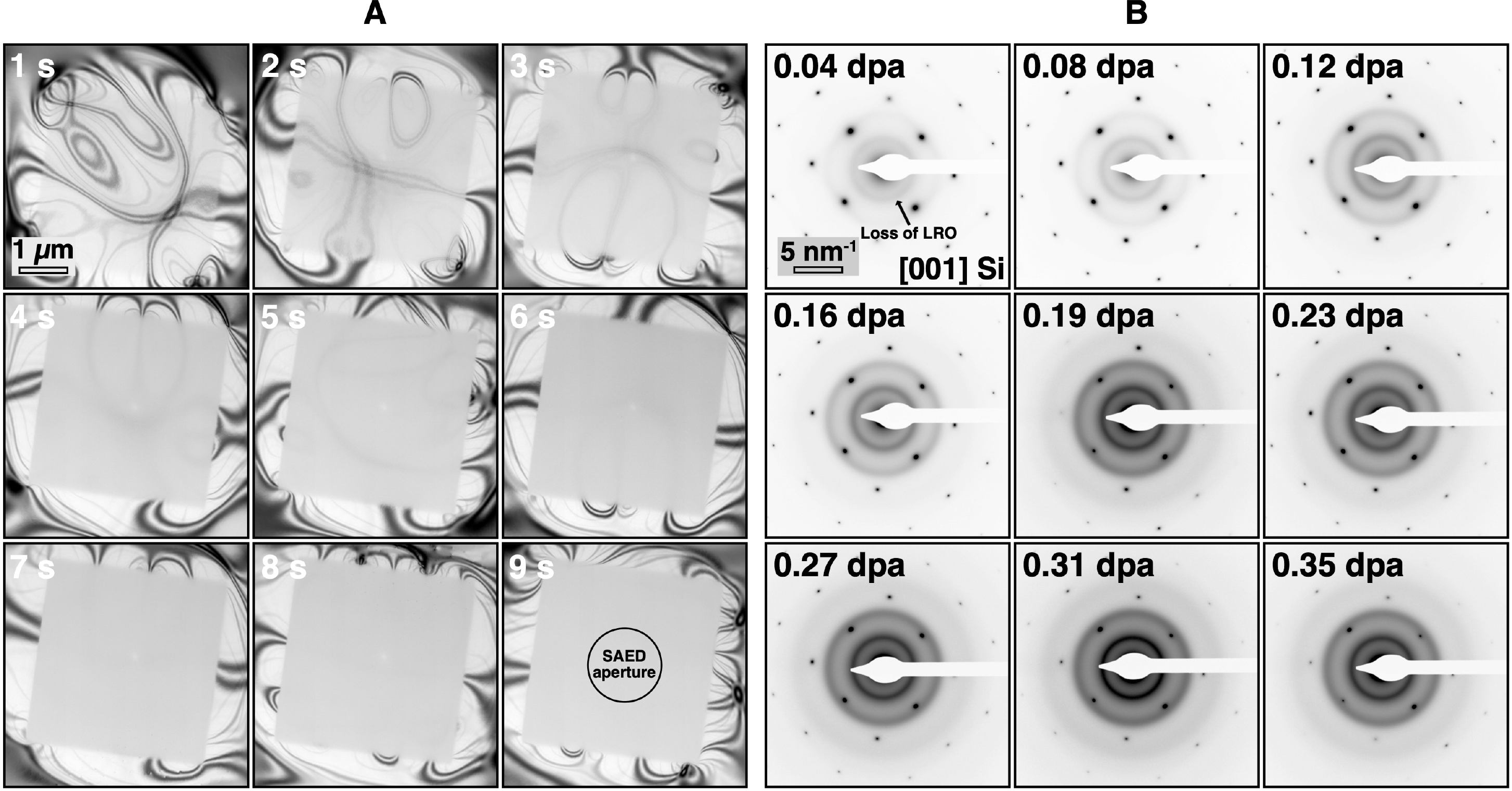}
		\caption{\textbf{Post-irradiation screening within a TEM} | The sample irradiated in the experiment shown in Fig. \ref{fig:03} was subjected to post-irradiation screening using a TEM in order to further evaluate the amorphization as a function of the irradiation dose. The set of BFTEM micrographs in \textbf{A} shows each site-specific irradiated area from 1 to 9 s. The respective SAED patterns taken along the [001] zone axis orientation are shown in \textbf{B}. As the dose increases from 0.04 to 0.35 dpa, clear and distinct diffusive rings became noticeable, which are attributed to the loss of crystallinity in the irradiated areas. The SAED patterns were recorded using an selected-area aperture that only covers a portion of the site-specific irradiated areas. As noted, complete amorphization was not attainable under the studied conditions (\textit{e.g.} irradiation time and dose rate) in this experiment. Note: the scale bars shown in \textbf{A} and \textbf{B} apply to all micrographs in their respective micrograph sets.}
		\label{fig:04}
	\end{figure}

	\noindent We now introduce a new irradiation mode to the radiation effects community: multi-area, multi-dose and site-specific irradiation. Bright-Field (BF), Dark-Field (DF) and High-Angle Annular Dark-Field (HAADF) micrographs in \textbf{Fig. \ref{fig:03}A-C}, respectively, show the microstructure of the pure Si sample taken before irradiation using the Scanning-Transmission Electron Microscopy (STEM) detector within the PFIB . The presence of bending contours -- characteristic of crystalline materials, caused by local variations in crystallographic planar orientations with respect to the incident electron beam -- is noted \cite{jacome2012advanced}. For multi-area, multi-dose ion irradiation, a series of nine 15$\times$15 $\mu$m$^{2}$ areas were exposed on the specimen. These areas were set to subjected to exposure times from 1 to 9 s, each area's parameters were specified using the microscope's standard software. A current of 10 pA was used, and under these irradiation conditions, the flux was 2.8$\times$10$^{13}$ ions$\cdot$cm$^{-2}\cdot$s$^{-1}$ and the average dose rate was estimated to be 0.035 dpa$\cdot$s$^{-1}$ for each irradiation area. The results of the ion irradiation within the PFIB using the conditions described above are presented in the BF, DF and HAADF micrographs in \textbf{Fig. \ref{fig:03}D-F}, respectively. A characteristic radiation effect on Si is observed in our experiment: as the fluence increases, the loss of bending contour contrast (as observed with three different imaging modes, \textit{i.e.} BF, DF and HAADF) within the square irradiation areas suggests the material is linearly losing its crystallinity, thus becoming amorphous, but only within the specified irradiated areas. The non-irradiated areas preserve their initial crystallinity. To the best of our knowledge, such an accurate tracking of Si amorphization using this new multi-area, multi-dose, and site-specific irradiation mode in a single specimen has never before been demonstrated. To unequivocally show that the amorphization phenomenon has ocurred, a post-irradiation characterization was performed within a TEM and the results are shown in \textbf{Fig. \ref{fig:04}A} and \textbf{Fig. \ref{fig:04}B}, respectively exhibiting BFTEM micrographs and Selected-Area Electron Diffraction (SAED) patterns. The appearance of amorphous diffusive rings as a function of dose is noted in the SAED patterns which were recorded from only within the irradiated square-areas using an appropriately sized SAED aperture. It is worth emphasizing that the first and most-intense amorphous ring present in the diffraction patterns is characteristic of loss of LRO in crystalline Si due to ion irradiation \cite{edmondson2009amorphization}. As noted in \textbf{Fig. \ref{fig:04}B}, full amorphization was still not observed up to an average dose of 0.35 dpa, corresponding to a fluence of 2.5$\times$10$^{13}$ ions$\cdot$cm$^{-2}$.
	
	\subsection*{Precise control of Si amorphization}
	\label{sec:results:precise}
	
	\noindent With regard to the feasibility of using the PFIB for ion-induced amorphization of pure Si electron-transparent specimens, one question remains to answered: how precisely can Si amorphization be performed? This question arises due to the non-uniform damage profile that 30 keV Xe$^{+}$ ion implantation causes on pure Si (\textbf{Fig. \ref{fig:02}B}). Conventionally, ion irradiation studies on this topic use a ``top hat'' damage profile to create a uniform distribution of defects along the specimen thickness, and often amorphization is observed as an uncontrolled phase transformation \cite{edmondson2009amorphization,edmondson2016situ}. For that, higher heavy-ion energies are required assuming an electron-transparent lamella. A second experiment was carried out to probe how precise and controlled one can control Si amorphization under irradiation within the PFIB. \textbf{Fig. \ref{fig:05}A} shows the pure Si specimen before irradiation. For this experiment, a single and larger area was drawn and the current was increased to 30 pA, resulting in a flux of 3.0$\times$10$^{13}$ ions$\cdot$cm$^{-2}\cdot$s$^{-1}$ and an average dose rate of 0.042 dpa$\cdot$s$^{-1}$. Micrographs after irradiation show, once again, that the irradiations induced amorphization in the target area (\textbf{Fig. \ref{fig:05}B}). Although full amorphization was still not observed at a dose of 0.63 dpa (4.5$\times$10$^{14}$ ions$\cdot$cm$^{-2}$), an almost fully amorphous Si region was produced as a result of PFIB irradiation and shown in the BFTEM micrograph and SAED pattern in \textbf{Fig. \ref{fig:05}C} and \textbf{Fig. \ref{fig:05}D}, respectively: the diffraction signals arising from the crystalline reflections of the \{022\} family of planes are still visible. This experiment demonstrates how powerful this methodology is as it allows a very precise control of irradiation dose right before the complete amorphization of Si. For comparison, Edmondson \textit{et al.} reported 0.88 dpa (2.0$\times$10$^{14}$ ions$\cdot$ cm$^{-2}$) as the critical dose for full amorphization of Si using 400 keV Xe$^{+}$ ion irradiation in a ``top hat'' damage profile \cite{edmondson2009amorphization}. The fact the almost full amorphization was observed in this paper at 0.63 dpa (4.5$\times$10$^{14}$ ions$\cdot$ cm$^{-2}$), the comparison with Edmondson's \textit{et al. in situ} TEM heavy-ion irradiation work is valid and suggests a precise control of Si amorphization under irradiation within the PFIB even considering a non-uniform damage profile posed by 30 keV Xe$^{+}$. It is worth emphasizing that higher fluences are required for amorphization of Si when using lower Xe acceleration energies.
	
	\section*{Discussion}
	\label{sec:discussion}
	
\begin{figure}	
		\centering	
		\includegraphics[scale=0.8]{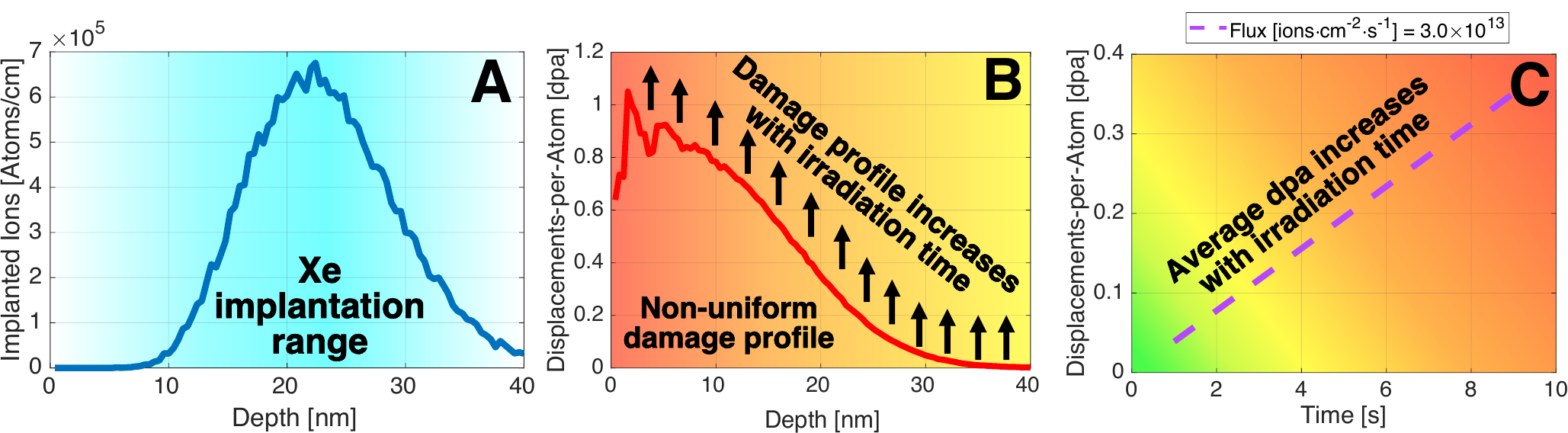}
		\caption{\textbf{Xe implantation and irradiation dose profiles calculated for pure Si} | SRIM2013-Pro calculations were performed by setting a 30 keV Xe$^{+}$ ion beam onto a Si target at normal incidence. An example case study shows that a small current of 30 pA over a 25$\times$25 $\mu$m$^{2}$ area generates an ion beam flux of 3.0$\times$10$^{13}$ ions$\cdot$cm$^{-2}\cdot$s$^{-1}$. The plot in \textbf{A} shows the implantation profile of Xe into Si demonstrating that this inert gas can be fully implanted within the Si TEM lamella. The (peak) dose profile as a function of irradiation depth is shown in \textbf{B}. The average dose over the 40 nm thick Si sample is shown as a function of time in plot \textbf{C}. For this case study, a dose rate of 0.042 dpa$\cdot$s$^{-1}$ was achieved. Under these studied conditions, an average dose of 100 dpa can be reached in around 40 min.}
		\label{fig:02}
	\end{figure}

\begin{figure}	
		\centering	
		\includegraphics[width=\textwidth]{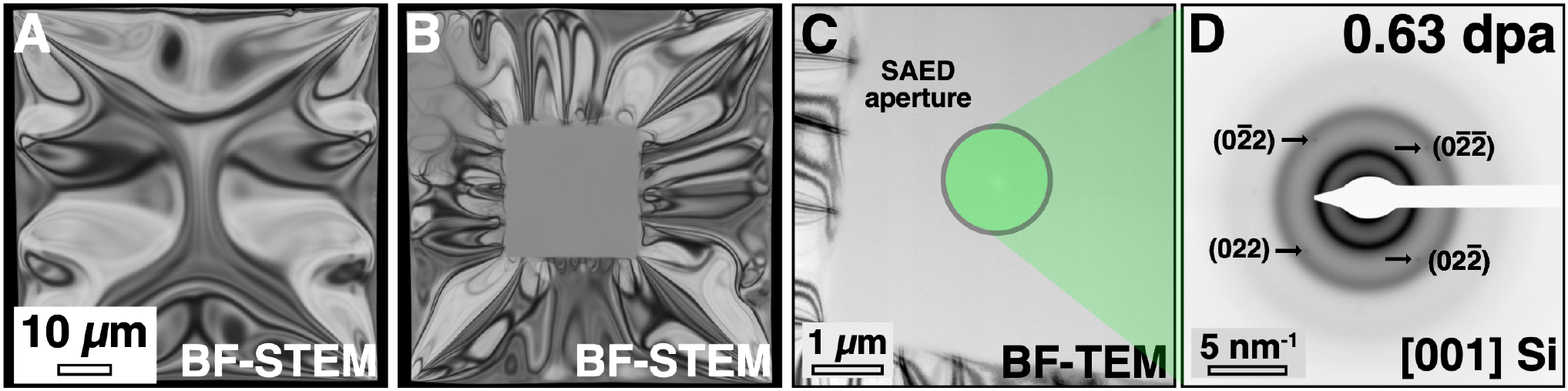}
		\caption{{\textbf{Probing controlled amorphization of Si with the PFIB} | For this experiment the dose rate was estimated to be 0.042 dpa$\cdot$s$^{-1}$, using a current of 30 pA and a single irradiation area of 15$\times$15 $\mu$m$^{2}$ (the irradiation conditions shown in Fig. \ref{fig:02}). BF micrograph taken before irradiation is shown in \textbf{A}. After irradiation, the amorphous area is noticeable in the center of BF micrograph in \textbf{B}. Post-irradiation characterization within the TEM using BFTEM (\textbf{C}) and SAED (\textbf{D}) indicates the sample almost fully amorphized at a dose of 0.63 dpa, which corresponds to a fluence of 4.5$\times$10$^{14}$ ions$\cdot$cm$^{-2}$. This experiment exemplifies how precise the ion beam modification of Si can be performed with the PFIB. Note: the scale bar in \textbf{A} also applies to micrographs \textbf{B}.}}
		\label{fig:05}
	\end{figure}

\begin{figure}	
		\centering	
		\includegraphics[width=\textwidth]{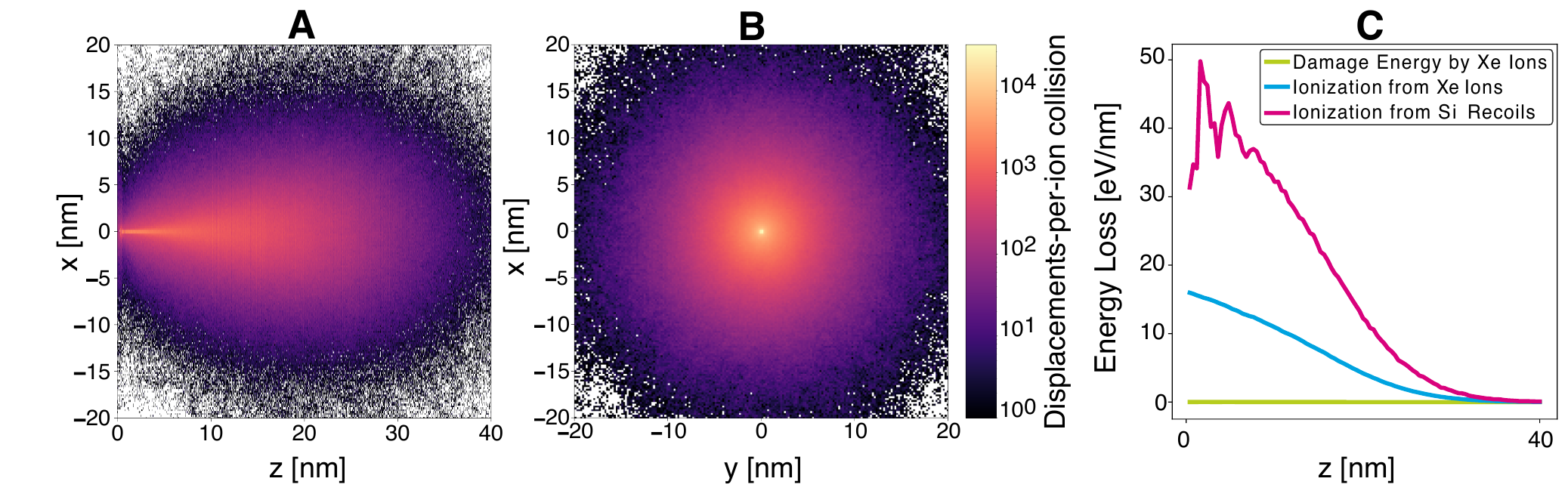}
		\caption{\textbf{Displacement damage cascade morphology and mechanisms of ion energy loss} | The pySRIM code \cite{ostrouchov2018pysrim} was used to estimate the cascade damage profiles of the 30 keV Xe$^{+}$ ion beam irradiation experiments into pure Si. For these calculations, 10$^{4}$ cascades were overlapped within a Si target with 40 nm as thickness. The plots in \textbf{A} and \textbf{B} show, respectively, the cascade morphology as a function of the implantation depth and the transverse view. Although the implantation profile presented in Fig. \ref{fig:02}A is not uniform over the irradiation depth, the distribution of defects generated by the displacement cascades can be considered homogenous over the thickness. Differences in ion energy loss mechanisms are shown in the plot in \textbf{C}.}
		\label{fig:06}
	\end{figure}

	\noindent To support these observations on the amorphization of Si with both non-uniform damage and implantation profiles caused by 30 keV Xe$^{+}$ ion irradiation, a series of calculations using pySRIM \cite{ostrouchov2018pysrim} were performed. As aforementioned, the Gibbons's heterogeneous model \cite{gibbons1972ion} is currently the most accepted hypothesis for the amorphization of semiconductors under ion irradiation. This theoretical model finds support and validation in numerous experiments and computational models. \cite{larubia1996defect,howe1981features,howe1987heavy,edmondson2009amorphization,edmondson2016situ,camara2019understanding} It assumes the creation of amorphous nano-zones in a material as a result of the overlapping  displacement cascades that are caused by each ion impact within the microstructure of the target material. Amorphization is therefore understood by the nucleation and interaction of these amorphous nano-zones as a function of the irradiation dose, resulting in complete loss of LRO. In order to better understand the amorphization phenomenon, an overlap of 10$^{4}$ displacement cascades caused by 30 keV Xe$^{+}$ ion impacts onto pure Si was simulated using pySRIM and the results are shown in \textbf{Fig. \ref{fig:06}A} and \textbf{Fig. \ref{fig:06}B}, respectively showing the cascade damage profiles as a function of specimen thickness and the transverse view. One can observe that while the 1D SRIM calculations exhibit non-uniform implantation and damage profiles (\textbf{Fig. \ref{fig:02}A-B}), the pySRIM analysis on cascade morphology and overlapping shows that damage over time can be uniformly distributed along the 40 nm thickness of the Si target. 
	
	These pySRIM calculations suggests that radiation damage starts accumulating in the first 20 nm layers of the material, then progressively with irradiation time/dose, the created amorphous nano-zones in this region will eventually overlap with the smaller zones in regions beyond 20 nm. Inevitably, such damage accumulation will result in complete amorphization of Si upon higher doses than the maximum herein studied (0.63 dpa). The pySRIM calculations can also be used to harness the physical mechanism by which Si amorphized. \textbf{Fig. \ref{fig:06}C} shows energy loss curves as a function of the target depth. These results show that ionization of the Si crystal -- \textit{i.e.} the energy transferred from Xe ions and Si recoils to electrons in the material --  is mainly dominated by Si recoils (magenta curve) after Xe ballistic ion collisions. The incident Xe ions also cause ionization of the target (blue curve), but to a lesser extent when compared to Si recoils. The damage energy deposition caused by Xe ions, where ionization from ions and recoils is not taken into consideration, is fairly small compared to the energy loss mechanisms involving recoil-induced ionization. Ion irradiation in this regime dominated by ionization indicates that the ``hot'' and disturbed electrons on the target material must dissipate this absorbed energy back into lattice. This causes a thermal spike via electron-phonon coupling \cite{zhang2017coupled}, which supports a core assumption of the Gibbons' model that a highly disturbed lattice region become amorphous after spikes caused by irradiation. Such a thermal spike can increase the local temperature up to a few thousand degrees in ultra-short times (picoseconds) \cite{stuchbery1999thermal}. Upon ultra-fast cooling, such ionized and disturbed Si nano-regions will be quenched-in amorphous. 
	
	We have hereby investigated the feasibility of using the PFIB as a new platform for fundamental radiation effects studies. The classical experiment on ion-induced amorphization of single-crystalline pure Si was selected as a proof-of-concept and the results indicate that irradiations within the PFIB  not only can induce amorphization, but allow highly controlled irradiations to be performed in a multi-area, multi-dose, and site specific irradiation mode which is unattainable by conventional ion acceleration/implantation systems. In addition, the PFIB  has high ion beam current and energy stabilities over extended periods of time resulting in a very low error spread in the current/flux, on the order of 1\%. It has also been demonstrated that very high dose rates can be achieved when using such a PFIB  irradiation methodology: dose rates in the order of 10$^{-2}$ dpa$\cdot$s$^{-1}$ are feasible using some of the smallest currents available for the PFIB (10-30 pA) without inducing any detectable sputtering effect. 
	
\begin{figure}	
		\centering	
		\includegraphics[width=\textwidth]{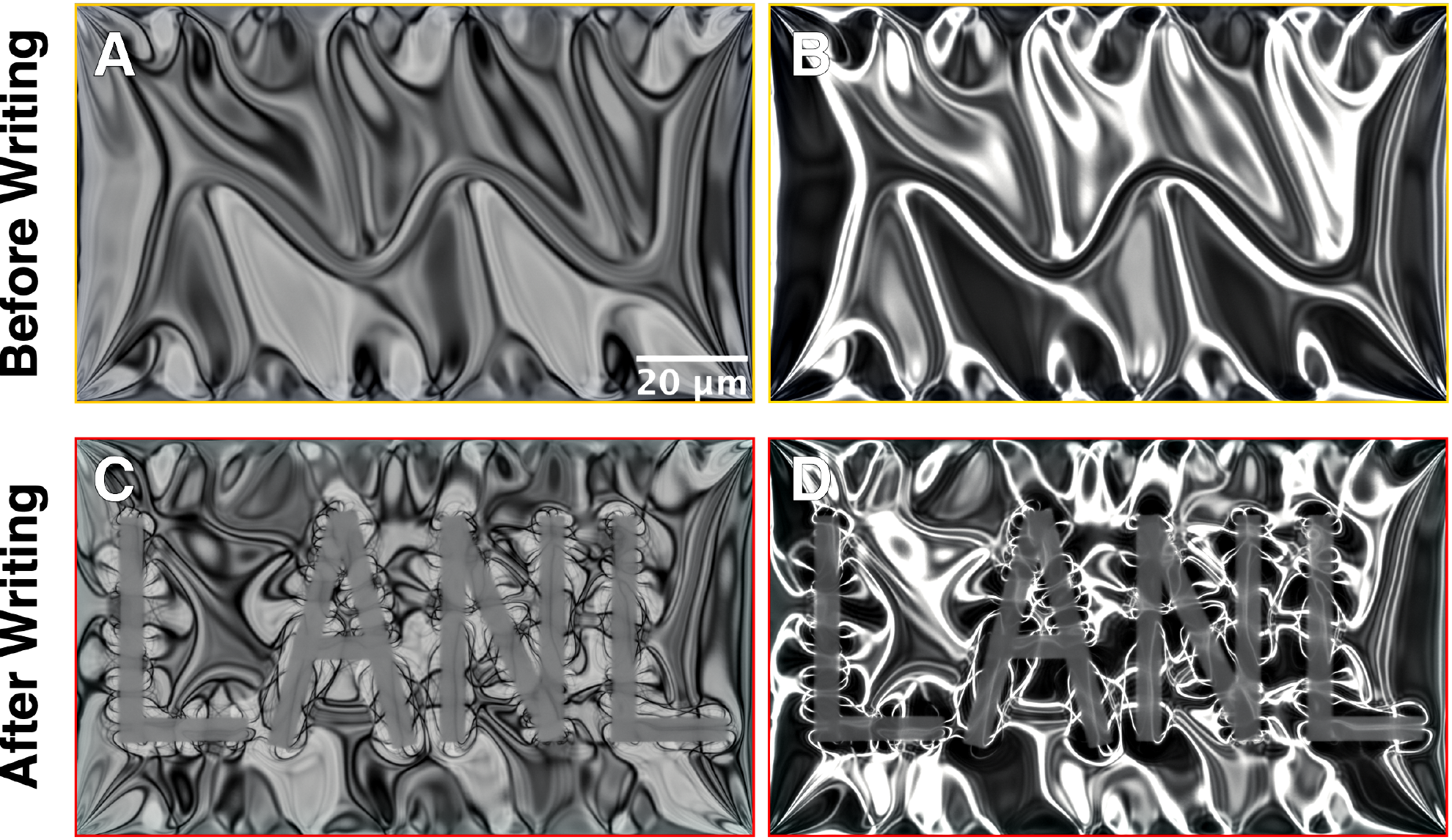}
		\caption{\textbf{Crystal-to-Amorphous Nanoscale Pen} | The careful control of the damage dose in site-specific areas suggests that ion irradiation within the PFIB arises as a new technique for nanofabrication and nanopatterning in solids. As an example, BF and DF-STEM micrographs \textbf{A, C} and \textbf{B, D} shows an electron-transparent pure single-crystal Si membrane before and after irradiation using a writing nanopattern built with the methodology described in this work. The printed areas correspond to amorphous Si whereas in the surrounding areas, Si remain fully crystalline. Given the differences in thermo, electronic and optical properties between amorphous and crystalline Si \cite{ovshinsky1978new,boyd1983laser}, the methodology described in this paper opens numerous possibilities for use the PFIB as a manufacturing tool of advanced nanodevices, harnessing the precise control of energetic particle irradiation provided which is inherent to PFIB. Notes: the scale bar in \textbf{A} applies to all micrographs in the figure and ``LANL'' is the acronym for Los Alamos National Laboratory.}
		\label{fig:07}
	\end{figure}

	An enormous number of new possibilities for future fundamental studies on radiation effects in solids are envisaged. Not exhaustively, Xe implantations of electron-transparent lamellae using the site-specific irradiation methodology introduced herein could lead to interesting studies on the nucleation and growth of inert gas bubbles in solids at a fundamental level, and Xe implantation and retention studies in potential nuclear fuel materials (\textit{e.g.} UN, UC, and UO2) at a technological level. Effects of interfaces such as grain boundaries, phase boundaries, or coatings, and their interaction with displacement-type defects, is just another example by which nano-scaled irradiation areas could shed light on the discovery of new phenomena. In a single specimen, multiple doses can be reached using the multi-area mode of irradiation. Given high achievable dose rates, irradiation in the PFIB  also supports testing of specimens under prototypic conditions up to the desired average dose levels for future nuclear reactors (100-500 dpa \cite{yvon2009structural,allen2010materials,zinkle2013materials,aguiar2020bringing}) in less than one hour. This technique can also be used to support irradiation testing using the recent idea of high-throughput synthesis of potential nuclear materials \cite{couet2022integrated}.	Nanodevice fabrication and nanopatterning is another field of research where the PFIB can contribute in future experimental ideas beyond the niche of materials at extremes. In a last experiment shown in \textbf{Fig. \ref{fig:07}A-D}, we demonstrate that in the case of pure Si, ion irradiation can be used to manufacture customized and controlled nanopatterns of amorphous regions within a crystalline matrix. To the best of our knowledge, such a ``Crystal-to-Amorphous Nanoscale Pen'' has never been demonstrated before in condensed matter physics. It is well known that amorphous and crystalline Si differ in thermo, electronic and optical properties \cite{ovshinsky1978new,boyd1983laser}, therefore, such PFIB  irradiation methods could lead to the fabrication of new nanodevices with irradiation-tailored properties. To finalize, the primary limitation of the methodology introduced in this paper is the limited acceleration energy of Xe within the PFIB, which is 30 keV, posing strict limits of both damage and implantation profiles below 50 nm for most materials.
	
	% Results
	
	% figure 01
	
	\begin{figure}	
		\centering	
		\includegraphics[width=\textwidth]{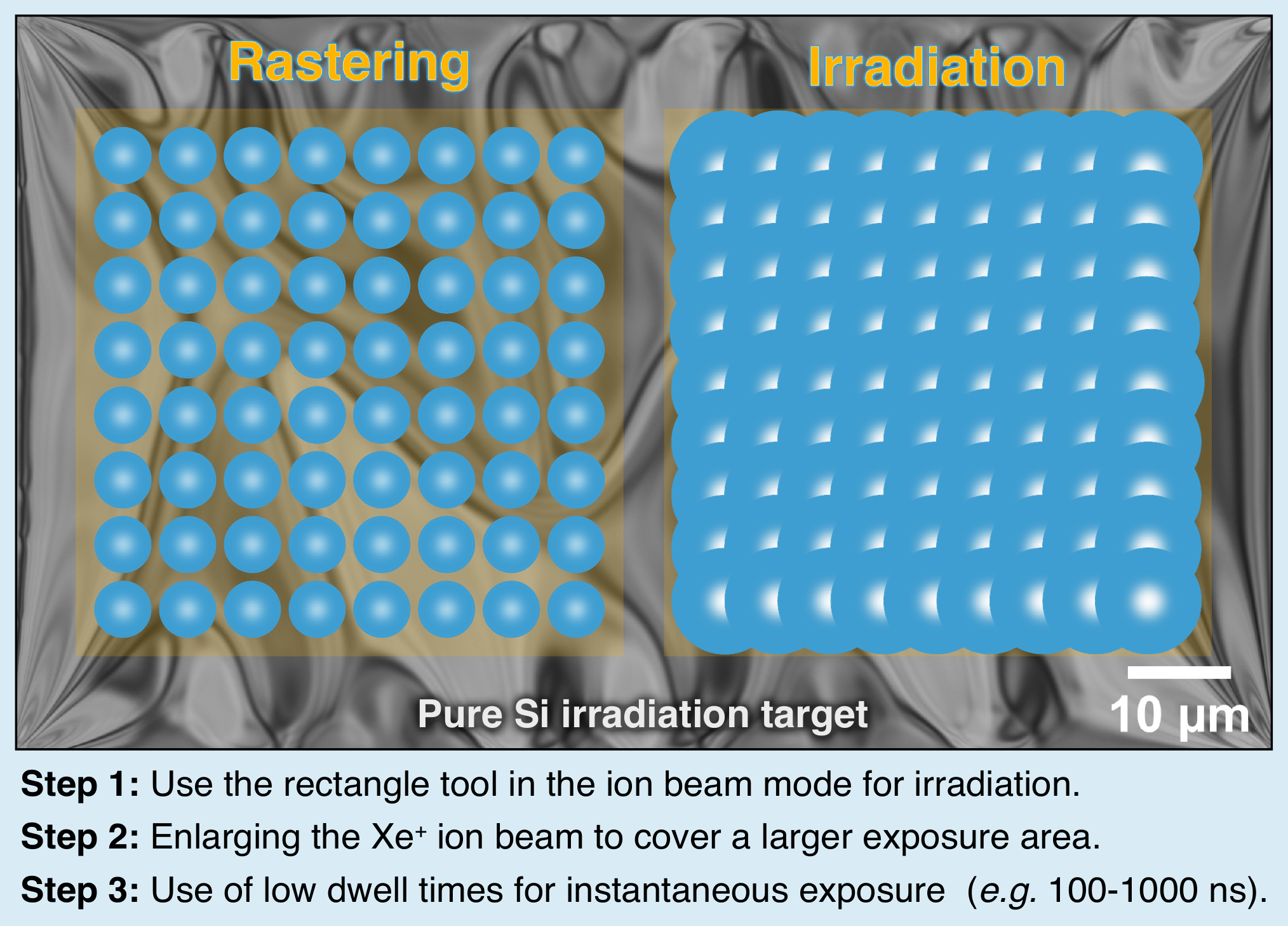}
		\caption{\textbf{Protocol for ion irradiation experiments using the PFIB} | The first step consists in selecting the rectangle tool within the ion beam mode. Under imaging conditions, the focused Xe ion beam is rastered over the specimen area. For irradiations, the second step requires enlarging the focused ion beam to approximately 50\% of its size to ensure that the ion beam will cover (almost) all the specimen area. The use of low dwell times (100-1000 ns) is the last step to establish a \textit{quasi}-instantaneous ion beam onto the target specimen over time. Following these steps, ion irradiations within the PFIB are conducted by tilting the specimen at a normal incidence angle with respect to the ion beam (\textit{i.e.} $52^{\circ}$), then selecting ion current (flux), irradiation area and time within the rectangle tool.}
		\label{fig:01}
	\end{figure}
	
	% figure 02
	
	% figure 03
	
	% figure 04
	
	% figure 05
	
	% figure 06
	
	% figure 07
	
	\section*{Materials and Methods}
	\label{sec:matmethods}
	
	\subsection*{Provenance of the pure and electron-transparent Si single-crystals TEM membranes}
	\label{sec:provenance}
	
	\noindent Pure and electron-transparent face-centered diamond cubic Si single-crystals were commercially acquired from the company SIMPore Precision Membrane Technology. The material consists of a TEM grid as a 100 $\mu$m thick frame containing eight 50$\times$50 $\mu$m$^2$ square windows and one 50$\times$100 $\mu$m$^2$ rectangular window. The pure Si specimens are found within these windows. Electron Energy-Loss Spectroscopy (EELS) was used in the TEM to estimate the thickness of the Si electron-transparent Si windows and it was estimated to be around 40 nm along the [001] zone axis. This method consists of estimating the ratio between the elastically scattered and the inelastic scattered electrons\cite{aronova2007quantification,egerton2008electron}.
	
	\subsection*{Ion irradiation within the PFIB }
	\label{sec:irradiations}
	
	\noindent A Thermo Fisher Helios G4 UXe dual-beam Scanning Electron Microscope (SEM) and Plasma Focused Ion Beam (PFIB) was used for all the experiments reported in this work. The materials were characterized before and after irradiation using the Scanning Transmission Electron Microscopy (STEM) detector within the PFIB. Samples were characterized in STEM-SEM using Bright-, Dark- and High-Angle Annular Dark-Field (BF, DF and HAADF) detectors. For the pre- and post-irradiation characterization within the PFIB, the electron beam energy was kept at 30 keV, which is significantly lower than the threshold for inducing recombination of point defects and/or causing other electron beam effects previously reported \cite{edmondson2009amorphization,edmondson2016situ}. Irradiations were carried out at a room temperature and at an angle of 52$^{\circ}$ with respect to the electron-beam, which results in normal incidence of the Xe$^{+}$ ion beam onto the specimen. The electron beam was not used for real-time \textit{in situ} STEM microstructural monitoring as the insertion of the STEM detector was not possible at the irradiation geometry, commercially available holders already on the market would enable such real-time SEM observation during ion beam irradiation. The irradiations were performed to homogeneously expose the irradiation area and dwell times were lowered to ensure instantaneous ion beam exposure (see Fig. \ref{fig:01}). Using this setup, both irradiation time and area can be precisely controlled using both the time exposed and the geometric configurations of the rectangle tool prior to irradiation. In this work, we have used low currents (10-100 pA) for irradiation as this was found to prevent sputtering of the irradiation areas. 
	
	Error estimation of the ion beam current is a major concern for this irradiation methodology. In fact, accurate ion beam currents are mandatory for commercial FIB systems. In order to estimate the error in the current within the PFIB, the ion beam is deflected in the condenser lens system into a Faraday cup built specifically for calibrating beam currents. A combination of the Xe gas leak-rate into the plasma system and the L1 lens (the first condenser lens) are adjusted until the 30 pA beam measures 30 pA$\pm$1\%. Subsequent beam currents are adjusted by varying the L1 lens only until the desired current is reached, also averaging an error around $\pm$1\%. These currents are checked and adjusted twice a year as part of PFIB  system's routine calibration and maintenance process. Thus, the ion flux can be accurately estimated using this methodology, by using error propagation, the error in the flux was also estimated to be in the order of 1\%. 
	
	The Stopping Range of Ions in Matter (SRIM) code was used to convert fluence to displacements-per-atom (dpa) \cite{ziegler2010srim}. For such a conversion, a method using the ``quick-damage'' calculation mode was selected following a procedure proposed by Stoller \textit{et al.} \cite{stoller2013use}. The parameters used for calculation were set as follows: (target) 40 nm as thickness, 25 eV for the bulk displacement energy \cite{holmstrom2008threshold,lefevre2009silicon,weber1998structure} and 2.33 g$\cdot$cm$^{-3}$ as density; (ion beam) Xe$^{+}$ ion specimen with 30 keV as acceleration energy and normal incidence. As an example, for an irradiation experiment within the PFIB using a current of 30 pA over a 25$\times$25 $\mu$m area, the estimated ion flux was around 3.0$\times$10$^{13}$ ions$\cdot$cm$^{-2}\cdot$s$^{-1}$, and a fluence of 4.5$\times$10$^{14}$ ions$\cdot$cm$^{-2}$ is reached in only 15 s of irradiation and corresponding to an average dose of 0.63 dpa over the 40 nm thickness (\textit{i.e.} an estimated dose rate of 0.042 dpa$\cdot$s$^{-1}$). Similarly, an average dose of 100 dpa is attainable in only 40 min of irradiation.
	
	The pySRIM code \cite{ostrouchov2018pysrim} -- which is a coupling of the SRIM code within the Python environment -- was used to both estimate the nuclear and electronic energy losses mechanisms and to carry out a cascade morphology overlapping study (using the ``full cascade'' calculation mode) of the irradiations on pure Si reported in this work.   
	
	\subsection*{Post-irradiation characterization within the TEM}
	\label{sec:postirradiation}
	
	\noindent After irradiations carried out using the PFIB, the samples were further analyzed using both a Thermo Fisher Tecnai TF30 and a Thermo Fisher Titan 80-300 Scanning/Transmission Electron Microscopes (S/TEM). Both BF-TEM and Selected Area Electron Diffraction (SAED) pattern modes were used for screening of the irradiated specimens.
	
	\section*{Acknowledgments}
	
	\noindent The Los Alamos National Laboratory (LANL) -- managed by Triad National Security LCC for the United States Department of Energy's (U.S. DOE) National Nuclear Security Administration (NNSA) -- provided research support to MAT via the Laboratory Directed Research and Development program under project number 20200689PRD2. This work has been carried out at the Electron Microscopy Laboratory at LANL. MAT is particularly grateful to all staff members of the Hydride and Fuels Research Laboratories at LANL for receiving him as a co-worker with patience and kindness.
	
	\section*{Data availability}
	\noindent All the data necessary to interpret the results presented in this research are in both main manuscript and supplemental information file.
	
	%\section*{References}
	\bibliographystyle{naturemag}%%elsarticle-num.bst for Elsevier journals
	\bibliography{bibdata}

\end{document}